\documentclass[twoside,twocolumn,9pt,a4paper]{article}
\usepackage{extsizes}
\usepackage[super,sort&compress,comma]{natbib} 
\usepackage[version=4]{mhchem}
\usepackage[left=1.5cm, right=1.5cm, top=1.785cm, bottom=2.0cm]{geometry}
\usepackage{balance}
\usepackage{times,mathptmx}
\usepackage{sectsty}
\usepackage{graphicx}
\usepackage{lastpage}
\usepackage{amsmath}
\usepackage{amssymb}
\usepackage[format=plain,justification=justified, singlelinecheck=false, font={stretch=1.125,small,sf}, labelfont=bf, labelsep=space]{caption}
\usepackage{float}
\usepackage{fancyhdr}
\usepackage{fnpos}
\usepackage[english]{babel}
\addto{\captionsenglish}{%
  
}
\usepackage{array}
\usepackage{droidsans}
\usepackage{charter}
\usepackage[T1]{fontenc}
\usepackage[usenames,dvipsnames]{xcolor}
\usepackage{setspace}
\usepackage[compact]{titlesec}
\usepackage{hyperref}

\usepackage{epstopdf} 

\definecolor{cream}{RGB}{222,217,201}

\begin{document}

\pagestyle{fancy}
\thispagestyle{plain}
\fancypagestyle{plain}{

\fancyhead[C]{\includegraphics[width=18.5cm]{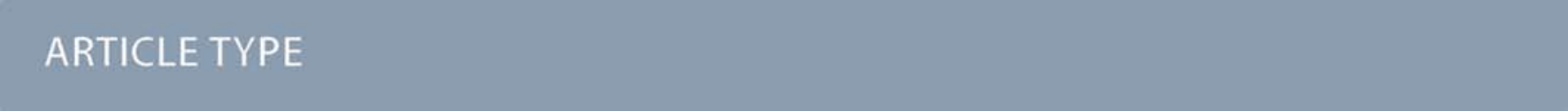}}
\fancyhead[L]{\hspace{0cm}\vspace{1.5cm}\includegraphics[height=30pt]{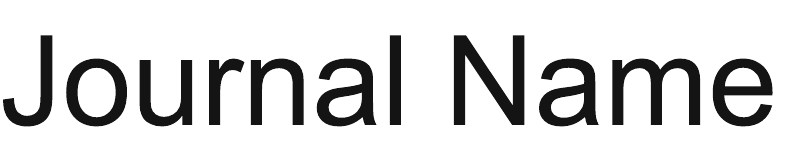}}
\fancyhead[R]{\hspace{0cm}\vspace{1.7cm}\includegraphics[height=55pt]{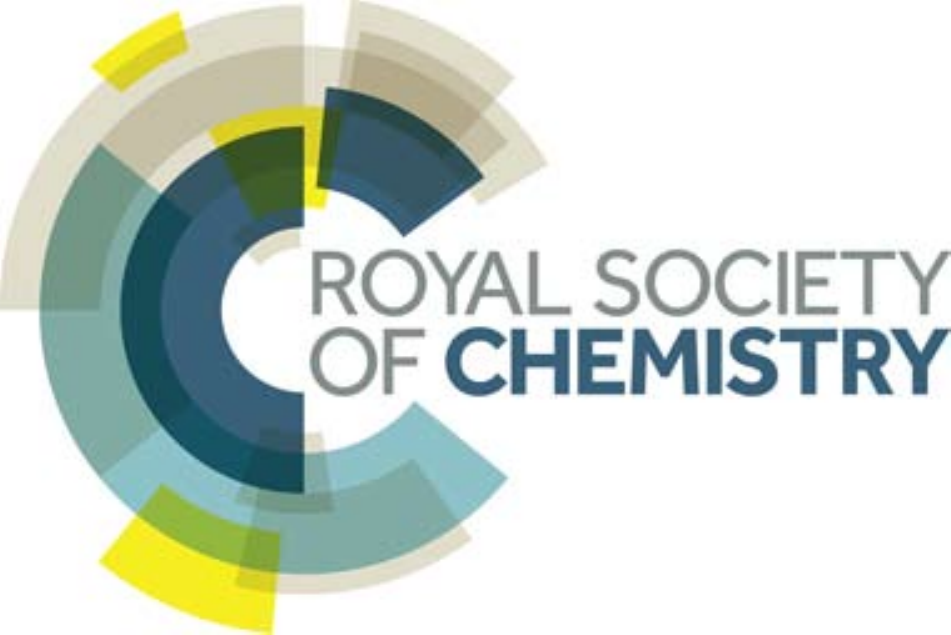}}
\renewcommand{\headrulewidth}{0pt}
}

\makeFNbottom
\makeatletter
\renewcommand\LARGE{\@setfontsize\LARGE{15pt}{17}}
\renewcommand\Large{\@setfontsize\Large{12pt}{14}}
\renewcommand\large{\@setfontsize\large{10pt}{12}}
\renewcommand\footnotesize{\@setfontsize\footnotesize{7pt}{10}}
\makeatother

\renewcommand{\thefootnote}{\fnsymbol{footnote}}
\renewcommand\footnoterule{\vspace*{1pt}%
\color{cream}\hrule width 3.5in height 0.4pt \color{black}\vspace*{5pt}} 
\setcounter{secnumdepth}{5}

\makeatletter 
\renewcommand\@biblabel[1]{#1}            
\renewcommand\@makefntext[1]%
{\noindent\makebox[0pt][r]{\@thefnmark\,}#1}
\makeatother 
\renewcommand{\figurename}{\small{Fig.}~}
\sectionfont{\sffamily\Large}
\subsectionfont{\normalsize}
\subsubsectionfont{\bf}
\setstretch{1.125} 
\setlength{\skip\footins}{0.8cm}
\setlength{\footnotesep}{0.25cm}
\setlength{\jot}{10pt}
\titlespacing*{\section}{0pt}{4pt}{4pt}
\titlespacing*{\subsection}{0pt}{15pt}{1pt}

\fancyfoot{}
\fancyfoot[LO,RE]{\vspace{-7.1pt}\includegraphics[height=9pt]{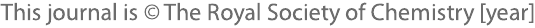}}
\fancyfoot[CO]{\vspace{-7.1pt}\hspace{13.2cm}\includegraphics{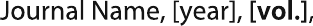}}
\fancyfoot[CE]{\vspace{-7.2pt}\hspace{-14.2cm}\includegraphics{RF}}
\fancyfoot[RO]{\footnotesize{\sffamily{1--\pageref{LastPage} ~\textbar  \hspace{2pt}\thepage}}}
\fancyfoot[LE]{\footnotesize{\sffamily{\thepage~\textbar\hspace{3.45cm} 1--\pageref{LastPage}}}}
\fancyhead{}
\renewcommand{\headrulewidth}{0pt} 
\renewcommand{\footrulewidth}{0pt}
\setlength{\arrayrulewidth}{1pt}
\setlength{\columnsep}{6.5mm}
\setlength\bibsep{1pt}

\makeatletter 
\newlength{\figrulesep} 
\setlength{\figrulesep}{0.5\textfloatsep} 

\newcommand{\topfigrule}{\vspace*{-1pt}%
\noindent{\color{cream}\rule[-\figrulesep]{\columnwidth}{1.5pt}} }

\newcommand{\botfigrule}{\vspace*{-2pt}%
\noindent{\color{cream}\rule[\figrulesep]{\columnwidth}{1.5pt}} }

\newcommand{\dblfigrule}{\vspace*{-1pt}%
\noindent{\color{cream}\rule[-\figrulesep]{\textwidth}{1.5pt}} }

\makeatother

\twocolumn[
  \begin{@twocolumnfalse}
\vspace{3cm}
\sffamily
\begin{tabular}{m{4.5cm} p{13.5cm} }

\includegraphics{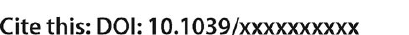} & \noindent\LARGE{\textbf{Particle-resolved lattice Boltzmann simulations of 3\nobreakdash-dimensional active turbulence$^\dag$}} \\
\vspace{0.3cm} & \vspace{0.3cm} \\

 & \noindent\large{D\'{o}ra~B\'{a}rdfalvy,\textit{$^{a}$} Henrik~Nordanger,\textit{$^{a}$} Cesare~Nardini,\textit{$^{b}$} Alexander~Morozov,\textit{$^{c}$} and Joakim~Stenhammar$^{\ast}$ \textit{$^{a}$}} \\

\includegraphics{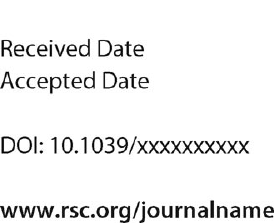} & \noindent\normalsize{Collective behaviour in suspensions of microswimmers is often dominated by the impact of long-ranged hydrodynamic interactions. These phenomena include active turbulence, where suspensions of pusher bacteria at sufficient densities exhibit large-scale, chaotic flows. To study this collective phenomenon, we use large-scale (up to $N=3\times 10^6$) particle-resolved lattice Boltzmann simulations of model microswimmers described by extended stresslets. Such system sizes enable us to obtain quantitative information about both the transition to active turbulence and characteristic features of the turbulent state itself. In the dilute limit, we test analytical predictions for a number of static and dynamic properties against our simulation results. For higher swimmer densities, where swimmer-swimmer interactions become significant, we numerically show that the length- and timescales of the turbulent flows increase steeply near the predicted finite-system transition density. 
} \\

\end{tabular}

 \end{@twocolumnfalse} \vspace{0.6cm}

]

\renewcommand*\rmdefault{bch}\normalfont\upshape
\rmfamily
\section*{}
\vspace{-1cm}


\footnotetext{\textit{$^{a}$~Division of Physical Chemistry, Lund University, Box 124, S-221 00 Lund, Sweden. E-mail: joakim.stenhammar@fkem1.lu.se}}
\footnotetext{\textit{$^{b}$~Service de Physique de l'\'{E}tat Condens\'{e}, CNRS UMR 3680, CEA-Saclay, 91191 Gif-sur-Yvette, France. }}
\footnotetext{\textit{$^{c}$~SUPA, School of Physics and Astronomy, The University of Edinburgh, James Clerk Maxwell Building, Peter Guthrie Tait Road, Edinburgh, EH9 3FD, United Kingdom. }}

\footnotetext{\dag~Electronic Supplementary Information (ESI) available: Videos showing the swimmers and the fluid in the disordered regime, transition regime and in the turbulent regime. See DOI: 10.1039/b000000x/}



\section{Introduction}

An archetypical example of active matter is a suspension of synthetic or biological particles that possess the ability to convert the energy extracted from their surroundings into self-propulsion.~\cite{Marchetti1,Bechinger1} This conversion of chemical energy into mechanical energy at the single-particle level and the resulting violation of detailed balance can result in rich displays of collective motion and dynamical self-assembly at larger lengthscales, such as strongly ordered bird flocks~\cite{Bialek1} and dynamical clustering in suspensions of active colloids.\cite{Palacci1} 

One of the most well-studied active systems is a suspension of bacteria or algae that swim by beating or rotating a collection of flagella.\cite{Lauga1,Cates1} At low densities, such systems show significantly enhanced diffusion of nonmotile particles compared to Brownian diffusion.\cite{Leptos1,Jepson1,Semeraro1,Kim1,Mino1} At higher concentrations, though still rather dilute, suspensions of rear-actuated (pusher) bacteria exhibit a complex collective behaviour known as \emph{active turbulence}, whereby the system starts exhibiting large-scale vortices and jets with higher fluid velocities than the velocity of the individual swimmer.~\cite{Creppy1,Cisneros1,Wensink1,Dunkel1} For front-actuated (puller) swimmers such as \emph{Chlamydomonas}, no such collective behaviour is observed in 3-dimensional suspensions,~\cite{Leptos1,Saintillan3} although instead a transition to a polar flocking state has been observed in simulations of puller stresslets confined to a 2-dimensional plane.~\cite{Menzel:MolPhys:2018,Menzel:JCP:2018} We also note that, in the case of \emph{squirmers}, which swim by an imposed slip flow along the spherical~\cite{Liverpool:PRE:2017,Pedley:IMA:2016,Alarcon:2013} or elongated~\cite{Gompper:SoftMatter:2018,Winkler:SoftMatter:2017} swimmer body, such a polar state is found for pullers also in 3 dimensions, while no sign of collective behaviour is found in the corresponding pusher suspensions.~\cite{Alarcon:2013,Liverpool:PRE:2017} While squirmers is a more appropriate model for ciliated organisms such as \emph{Paramecium}, these different collective behaviours highlight that the aspect ratio and seemingly subtle differences in the near-field flows can have large impacts on the non-equilibrium steady states. 

The transition to active turbulence in pusher suspensions has been described as a hydrodynamic instability induced by the mutual reorientation of swimmers due to the long-ranged stresslet flow fields.~\cite{Marchetti1,Stenhammar1,Yeomans1,Saintillan1,Simha1,Subramanian1,Wolgemuth1,Koch:AnnuRev:2011} Much of the theoretical understanding of active turbulence is based on continuum theories that describe the active suspension using effective equations of motion for the order parameter fields. These are typically derived from either a kinetic theory of a set of stresslet swimmers,\cite{Saintillan3,Wolgemuth1,Subramanian1,Stenhammar1,Marchetti1} or through the modification of the equations describing nematic liquid crystals through additional active stresses, resulting in so called active nematics models.~\cite{Doostmohammadi:NatComm:2018,Thampi1,Ramaswamy1} For the former class of models, the microscopic parameters describing the transition to active turbulence can be determined through stability analysis of the linearised equations. For an unbounded suspension, the most unstable mode is the $k = 0$ one, and the ensuing analysis leads to the following prediction for the critical pusher number density $n_c$ required for collective motion:\cite{Saintillan1,Subramanian1,Stenhammar1,Saintillan4,Hohenegger1}
\begin{equation}\label{rho_c}
n_c = 5\lambda / \kappa.
\end{equation}
Here, $\lambda$ is the tumbling frequency by which individual swimmers randomise their swimming direction and $\kappa$ is the stresslet magnitude, defined below. In order to probe the properties of the turbulent state itself, the linear theory is no longer accurate, and one instead needs to numerically integrate the equations of motion.~\cite{Saintillan4,Saintillan:SoftMatter:2017,Ezhilan:PoF:2013,Doostmohammadi:NatComm:2018} This approach gives access to the full nonlinear behaviour of the turbulent state, although still resting on the approximations underlying the continuum equations. Particle-based simulation studies of active turbulence are more limited in number, partially due to the computational challenges of simulating large collections of hydrodynamically interacting particles over extended length- and timescales. These difficulties have so far prohibited a quantitative verification of the analytical predictions, such as that of the transition density. Hernandez-Ortiz \emph{et al.}~\cite{Underhill1,Hernandez1} and Lushi and Peskin~\cite{Lushi1} employed a model that describes each pusher swimmer as two connected spheres with a pair of embedded point forces, and observed a transition to a collectively flowing state resembling active turbulence. They characterised the coherent flows by calculating the properties of the fluid flows and the enhanced diffusion of tracer particles. Subsequently, Saintillan and Shelley~\cite{Saintillan2} and Krishnamurthy and Subramanian~\cite{Krishnamurthy1} observed similar collective properties in suspensions containing up to $N = 3\times10^4$ slender rod-like particles with pusher flow-fields in 3 dimensions. 

In this paper, we significantly extend the above studies to system sizes reaching $N > 10^6$ microswimmers, which enables us to quantitatively study the properties of active turbulence with high numerical accuracy over extended length- and timescales. The simulations are based on an implementation of the lattice Boltzmann (LB) equation~\cite{Nash1} that allows us to describe each swimmer as a pair of point forces acting on the surrounding fluid, thus ignoring the effect of near-field hydrodynamics and excluded volume interactions, which can be justified by the relatively low number densities needed to reach the turbulent state. We show that the length- and timescales of the chaotic flows increase sharply at the transition to turbulence, before decreasing towards a plateau value in the turbulent state. The swimmer density where this increase is observed is in agreement with the theoretical prediction from kinetic theory, as long as we take into account the finite-size effects from using a finite box with periodic boundary conditions. We also show that studying these phenomena require very large system sizes: our results suggest that linear box dimensions of at least $\sim 100$ times the swimmer length are required to eliminate finite-size effects. These large systems furthermore allow us to study the statistics of swimmer-swimmer correlations with unprecedented accuracy: the spatial two-body correlation functions show a transition to a state of strong orientational order, induced purely by far-field hydrodynamic interactions, but with no signs of significant density inhomogeneities. Our results provide a thorough characterisation of the hydrodynamically induced collective motion in pusher suspensions which should pave the way both for further experimental efforts and for testing analytical descriptions of active turbulence. 

\section{Model and Methods}

\begin{figure}[t]
 \centering
  \includegraphics[height=3cm]{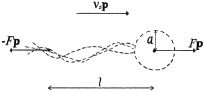}
  \caption{Schematic image of the pusher model, where $F$ is the force, $l$ is the swimmer length, $v_s$ the swimming speed, $a$ the effective body radius and $\mathbf{p}$ is the orientation of the swimmer.}
  \label{fig:swimmer}
\end{figure}

We consider a suspension of $N$ swimmers represented by extended force dipoles (stresslets) moving in a three-dimensional fluid with periodic boundaries. The body and flagella exert two equal and opposite forces $\pm F \mathbf{p}$, where $\mathbf{p}$ is the orientation of the swimmer, separated by a length $l$ on the fluid as shown in Fig.~\ref{fig:swimmer}. The swimmer is characterised by a dipole strength $\kappa = \pm Fl/ \mu$ where $\mu$ is the dynamic viscosity of the fluid, with $\kappa > 0$ representing pushers and $\kappa < 0$ pullers. 

The position $\mathbf{r}$ and orientation $\mathbf{p}$ of each swimmer evolves according to the following equations of motion:~\cite{Saintillan:PoF:2008,Stenhammar1}
\begin{eqnarray}
\label{rdot} &\dot{\mathbf{r}}=v_s\mathbf{p} + \mathbf{U(r)}, \\
\label{pdot} &\dot{\mathbf{p}}=(\mathbb{I}-\mathbf{p}\mathbf{p})\cdot (\nabla\mathbf{U})\cdot\mathbf{p} \approx (\mathbb{I}-\mathbf{p}\mathbf{p})\cdot \displaystyle{\frac{\mathbf{U(r)-U(r-\mathbf{p}}l)}{l}}.
\end{eqnarray}
Here, $\mathbf{U(r)}$ is the fluid velocity evaluated at the body position, $v_s$ is the constant swimming velocity of the individual swimmer, and $\mathbb{I}$ is the unit tensor. Equation \eqref{pdot} is the (discretised version of) Jeffery's equation for infinite aspect ratio ($\beta = 1$); we checked our results also using a finite aspect ratio ($\beta < 1$), which yielded only small shifts for realistic values of $\beta$. In addition to their reorientation as a result of the hydrodynamic interactions described by Eq. \eqref{pdot}, swimmers also undergo Poisson-distributed random reorientations with an average frequency $\lambda$. (Note that this mechanism cannot be cast as a continuous-time differential equation as that in Eq.~\eqref{pdot}.~\cite{Subramanian1}) This run-and-tumble motion results in a random walk with a persistence length $v_s/\lambda$. 

Note that, in our model, the symmetry breaking in the single swimmer dynamics is \emph{only} created by the self-propulsion term in the equation of motion \eqref{rdot}: the single-swimmer flow-field is fully fore-aft symmetric, as there is no surface describing the swimmer body. However, an effective body radius $a$ for the swimmer geometry in Fig.~\ref{fig:swimmer} can be calculated through the following relation between $F, v_s$, and $a$:\cite{Stenhammar1}
\begin{equation}\label{swimmer_radius}
v_s=\frac{F}{6\pi\mu a} \left( 1-\frac{3a}{2l} \right).
\end{equation}
Here, the second term in brackets constitutes the leading-order correction to the Stokes-Einstein relation for the spherical swimmer body due to the disturbance flow created by the flagellar Stokeslet.

\begin{figure*}[t]
 \centering
 \includegraphics[height=5.5cm]{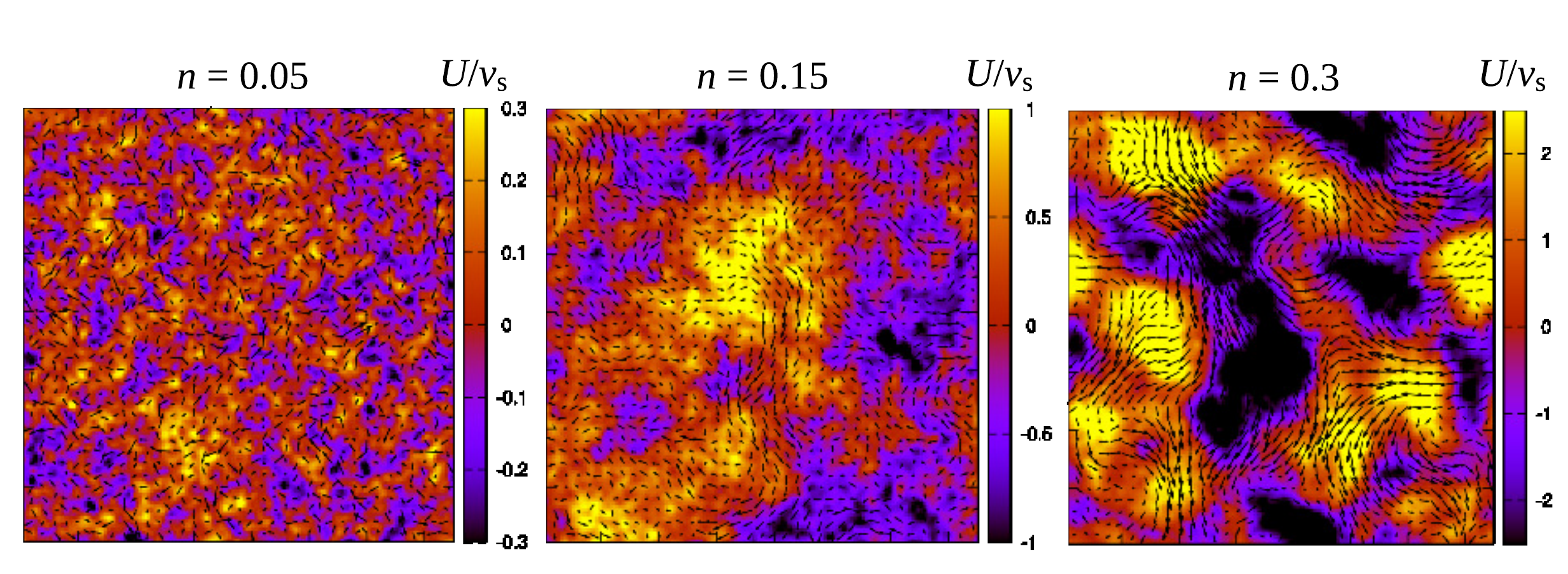}
 \caption{Snapshots of the fluid velocity field in a 2-dimensional lattice plane of the 3-dimensional simulation box ($L = 150$) at densities before the transition to turbulence ($n=0.05$), close to the transition ($n=0.15$), and in the turbulent regime ($n=0.3$). Vectors show the velocity field in the $xy$-plane and the colours indicate its $z$-component. The length of the velocity vectors have been rescaled for clarity. See also the corresponding videos available as ESI\dag. }
 \label{fig:snapshots}
\end{figure*}

To simulate the hydrodynamic interaction between microswimmers, we used a D3Q15 BGK lattice Boltzmann (LB) method based on the point-force LB implementation of Nash \emph{et al.}~\cite{Nash1,Nash2} The key ingredient in this method is an algorithm for interpolating forces and velocities between the fluid and the off-lattice swimmers that employs the regularised version of the $\delta$ function due to Peskin.~\cite{Nash1,Peskin:2002} Since this function has a compact support of 2 lattice units, it effectively gives non-singular flow-fields that are distributed over distances comparable to the LB lattice spacing. Importantly, the method enables large-scale simulations of up to $N \sim 10^6$ microswimmers at biologically relevant densities. In the simulations, cubic box sizes $L^3$ ranging from $(10)^3$ to $(210)^3$ lattice sites were employed. In terms of LB units (where $\Delta L = \Delta t = 1$), we used the parameters $v_s=10^{-3}$, $F=1.57 \times 10^{-3}$, $l = 1$, $\lambda = 2 \times 10^{-4}$, and $\mu=1/6$, with the latter value corresponding to the fluid relaxing to local equilibrium on each timestep. Each simulation was run for $2 \times 10^5$ timesteps, apart from in the transition region where longer simulations were necessary due to the slow dynamics. A typical simulation with $L = 100$ took $\sim 48$ hours on a single CPU, while the largest simulations ($L = 210$) took $\sim 1$ week in parallel on 5 CPUs.

All results will be presented in terms of the swimmer length $l$ and the time scale $l / v_s$.\footnote{An alternative way of non-dimensionalising the time units is to use the characteristic tumbling time $\lambda^{-1}$, which would be the relevant timescale in a suspension of non-swimming stresslets (``shakers'').\cite{Stenhammar1}} The relevant dimensionless numbers of the system are (\emph{i}) the single-swimmer Reynolds number $\mathrm{Re}_s \equiv \rho_{\mathrm{f}} v_s l / \mu$, where $\rho_{\mathrm{f}}$ is the fluid density (set to unity in our LB simulations), and (\emph{ii}) the nondimensional stresslet strength $\kappa_n \equiv \kappa / (l^2 v_s) = F / (\mu l v_s)$. The above parameter values yield $\mathrm{Re}_s = 6 \times 10^{-3}$, which is well below the Stokes-flow limit,\cite{deGraaf1} and  $\kappa_n \approx 9.4$. The latter value can be compared with the corresponding value for \emph{E. coli} for which $v_s \approx 22$ $\mathrm{\mu m s^{-1}}$, $F = 0.42$ pN, and $l = 1.9$ $\mathrm{\mu m}$,~\cite{Drescher1} yielding $\kappa_n \approx 11.2$. Furthermore, an approximate volume fraction based on the spherical swimmer body can be calculated with the aid of Eq. \eqref{swimmer_radius} as $\phi = (4\pi/3) a^3 n$, where $n = N/L^3$ is the swimmer number density; with our parameters, Eq. \eqref{swimmer_radius} gives $a \approx 0.3$. Using this estimate, the observed critical volume fraction needed for collective motion is $\phi_c \approx 0.02$, which is clearly within the range of validity of our far-field hydrodynamic model. While the highest densities considered in this paper ($\phi \approx 0.055$) are large enough that near-field effects and other specific interactions would start to become significant, we would like to highlight that these strongly turbulent flows can also be achieved at small densities if the dipolar strength $\kappa$ is large enough, in accordance with Eq.~\eqref{rho_c}. A relatively high-density suspension of weak dipoles can thus be viewed as a proxy for the turbulent properties of a dilute suspension of strong dipoles.

\section{Analytical expressions for noninteracting swimmers}

In this section, we will derive analytical results for the statistical and dynamical properties of a collection of \emph{noninteracting} microswimmers, where we set the terms containing $\mathbf{U}$ in Eqs. \eqref{rdot} and \eqref{pdot} to zero. The swimmers still exert pairs of equal and opposite forces on the fluid, and generate long-ranged flow fields. Note that, in the non-interacting case, pushers and pullers are equivalent. In this limit, many average properties of the system can be obtained analytically with relative ease, since the swimmers are then statistically independent. In particular, we calculate the spatial and temporal correlation functions of the fluid, the fluid velocity variance, and the associated Fourier-space energy spectrum. These predictions will then be compared in the following sections to the corresponding LB simulations of non-interacting swimmers. The corresponding description taking into account swimmer-swimmer interactions is still achievable~\cite{Stenhammar1} but considerably more complicated; we therefore postpone the full analytical treatment of interacting swimmers to a forthcoming separate study.

Following Cortez \emph{et al.},~\cite{Cortez2005} we start from the expression for the regularised flow field from a point force with magnitude $F$:
\begin{equation}
u_i({\bf r}) = \frac{(r^2 +2\epsilon^2)\delta_{ij} + r_i r_j}{\left(r^2 + \epsilon^2 \right)^{3/2}}\frac{F_j}{8\pi\mu}.
\end{equation}
Here, $\epsilon$ is a factor describing the distance over which the regularisation acts and $r = |\mathbf{r}|$; in the limit $\epsilon \rightarrow 0$ this expression reduces to the usual Stokeslet. Its Fourier transform is given by
\begin{equation}
\hat u_i({\bf k}) = \left(\delta_{ij} - \hat{k}_i \hat{k}_j  \right) F_j \frac{\epsilon^2 K_2(k \epsilon)}{2\mu},
\end{equation}
where $K_2$ is the modified Bessel function of the second kind, and $\hat k_i = k_i/k$, where $k=|{\bf k}|$. The velocity field of an extended dipole is constructed from this expression by placing two point forces of equal magnitude $F$ and opposite orientations at $\mathbf{r}_0$ and $\mathbf{r}_0 + l\mathbf{p}$, where $l$ is the dipolar length and $\bf p$ its orientation. The corresponding real-space velocity field of a single swimmer is thus
\begin{align}
&u_i(\mathbf{r};\mathbf{r}_0,\mathbf{p}) = \frac{1}{(2\pi)^3} \int d{\bf k}\, e^{i {\bf k}\cdot ( {\bf r} - {\bf r}_0 )}
\left[ e^{- i{\bf k} \cdot l{\bf p}} - 1 \right] \nonumber \\
& \qquad\qquad\qquad\qquad\qquad \times \left( \delta_{ij} - \hat{k}_i \hat{k}_j  \right) p_j \frac{F \epsilon^2 K_2(k \epsilon)}{2\mu}.
\end{align}
The total velocity field at a position ${\bf r}$ created by a suspension of $N$ non-interacting swimmers with instantaneous positions ${\bf r}_i$ and orientations ${\bf p}_i$, $i=1\dots N$ is, then,
\begin{equation}
\mathbf{U}({\bf r}) = \sum_{i=1}^N \mathbf{u}({\bf r};{\bf r}_i,{\bf p}_i).
\end{equation}

\subsection{Temporal correlations}

First we consider the velocity-velocity autocorrelation function $c(t) = \langle {\bf U}(0) \cdot {\bf U}(t)\rangle$, which is given by the following ensemble average 
\begin{align}
& c(t) = \frac{e^{-\lambda t}}{V}\int d{\bf r} \frac{1}{V^N}\int d{\bf r}_1 \dots d{\bf r}_N 
\frac{1}{(4\pi)^N}\int d{\bf p}_1 \dots d{\bf p}_N \nonumber \\
&\qquad\qquad \times\left( \sum_{j=1}^N {\bf u}({\bf r};{\bf r}_j,{\bf p}_j)\right)
\cdot \left( \sum_{k=1}^N {\bf u}({\bf r};{\bf r}_k + v_s t {\bf p}_k,{\bf p}_k)\right).
\end{align}
Here, we used the fact that the time-dependence of the velocity field only arises through changes of the swimmer positions; the factor $e^{-\lambda t}$ accounts for independent decorrelation events due to tumbling, and $v_s$ is the swimming speed. Since the swimmers are statistically independent, only 'self-correlations' contribute to the average, yielding
\begin{align}
c(t) = \frac{n}{2\pi^2} \int_0^\infty dk k^2 I_1(k) \left(\frac{F \epsilon^2 K_2(k \epsilon)}{2\mu}\right)^2,
\end{align}
where
\begin{align}
I_1(k) = \int_{-1}^{1} dx \, e^{-\lambda t-i k v_s t x}\left[ 1-\cos{(k l x)}\right] \left(1-x^2\right).
\end{align}
Here, as before, $n=N/V$, is the swimmer number density.
Explicit integration gives
\begin{align}\label{c_t_final}
c(t_n) = e^{-\lambda_n t_n} \frac{g_1(t_n+1)-2g_1(t_n)+g_1(t_n-1)}{2g_1(1) - \frac{5\pi\Delta^2}{8}},
\end{align}
where
\begin{align}
g_1(x) = \frac{ 1+x^2  \Delta^2}{x^2} \mathbb{E}\left(-\frac{1}{4} x^2 \Delta^2 \right) - \frac{ 1+\frac{1}{2}x^2  \Delta^2}{x^2}\mathbb{K}\left(-\frac{1}{4} x^2 \Delta^2 \right),
\end{align}
with $\mathbb{E}(x)$ and $\mathbb{K}(x)$ being the complete elliptic integrals of the first and second kind, respectively. We have furthermore introduced the dimensionless time $t_n = t v_s /l$, tumbling rate $\lambda_n = \lambda l/v_s$, and regularisation parameter $\Delta=l/\epsilon$, and we have normalised $c(t_n)$ so that $c(0)=1$.
\subsection{Spatial correlations}
Similar to the temporal correlation function, the spatial velocity-velocity correlation function $c(R) = \langle {\bf U}({\bf 0}) \cdot {\bf U}({\bf R})\rangle$ is given by the following ensemble average
\begin{align}
&c(R) = \frac{1}{V}\int d{\bf r} \frac{1}{V^N}\int d{\bf r}_1 \dots d{\bf r}_N \frac{1}{(4\pi)^N}\int d{\bf p}_1 \dots d{\bf p}_N \nonumber \\
& \qquad\qquad\qquad\qquad\times {\bf U}({\bf r}) \cdot {\bf U}({\bf r}+{\bf R}).
\end{align}
Again, keeping only the 'self-correlation' contributions, we obtain
\begin{align}
&c(R) = \frac{4 n}{(2 \pi)^3} \nonumber \\
&\qquad \times \int d{\bf k} \, e^{- i {\bf k}\cdot {\bf R}} \left[\frac{1}{3} + \frac{\cos k l}{(k l)^2} - \frac{\sin k l}{(k l)^3}\right] \left(\frac{F \epsilon^2 K_2(k \epsilon)}{2\mu}\right)^2.
\label{cR}
\end{align}
Direct evaluation of this integral yields
\begin{align}\label{c_R_final}
&c(R_n) = \frac{1}{c_0 R_n} \left[ g_2(R_n+1) -\frac{10 \Delta^2}{4+\Delta^2 R_n^2} g_2(R_n) + g_2(R_n-1) \right. \nonumber \\
& \qquad\qquad\quad \left. g_3(R_n+1) -\frac{10 \Delta^2}{R_n\left(4+\Delta^2 R_n^2\right)} g_3(R_n) - g_3(R_n-1)
\right],
\end{align}
where 
\begin{align}
&g_2(x) = 10\,x\,\Delta^2 \left[ 2\left(5+ x^2  \Delta^2\right) \mathbb{E}\left(-\frac{1}{4} x^2 \Delta^2 \right) \right. \nonumber \\
& \qquad\qquad\qquad\qquad \left. - \left(4+ x^2  \Delta^2\right) \mathbb{K}\left(-\frac{1}{4} x^2 \Delta^2 \right)
\right], \\
&g_3(x) = 2\left( x^4 \Delta^4 + 2 x^2 \Delta^2 -8 \right) \mathbb{K}\left(-\frac{1}{4} x^2 \Delta^2 \right) \nonumber\\
& \qquad\qquad -4\left( x^4 \Delta^4 + 9 x^2 \Delta^2 - 4 \right) \mathbb{E}\left(-\frac{1}{4} x^2 \Delta^2 \right),
\end{align}
and
\begin{align}
&c_0 = 15 \Delta^2 \left[8(1+\Delta^2)\mathbb{E}\left(-\frac{1}{4}\Delta^2 \right) \right. \nonumber \\
& \qquad \qquad\qquad \qquad \left. - 4(2+\Delta^2)\mathbb{K}\left(-\frac{1}{4}\Delta^2 \right) -\frac{5\pi}{2} \Delta^2\right].
\end{align}
Here, $R_n = R/l$, and we have again normalised $c(R_n)$ such that $c(0)=1$. For large $R_n$, $c(R_n)\sim R_n^{-1}$, in agreement with the results of Zaid \emph{et al.}~\cite{Zaid2011}

\subsection{Velocity variance}

The velocity variance $\langle U^2 \rangle$ can be obtained from Eq.\eqref{cR} by setting $R=0$. The result depends on $c_0$ and reads
\begin{align}
\langle U^2 \rangle  = \frac{\kappa^2 n}{480 \pi} \frac{c_0}{\epsilon \Delta^6}.
\end{align}
By combining the large- and small-$\Delta$ asymptotics of this expression, we obtain the following uniform approximation, which interpolates well between the two regimes and is significantly easier to use than the full expression: 
\begin{equation}\label{eq:Usq}
\langle U^2 \rangle \approx \frac{21 n \kappa^2}{8 \left(21 \pi l + 256 \epsilon \right)}.
\end{equation}

\subsection{Energy spectrum}

Eq.\eqref{cR} can be interpreted as a Fourier transform of the velocity-velocity spatial correlation function, and we therefore can identify (after the substitution ${\bf k}\rightarrow {-\bf k}$ due to our definition of the Fourier transform)
\begin{equation}
\langle \hat{\bf U}({\bf k}) \cdot \hat{\bf U}(-{\bf k})\rangle =
4 n \left[\frac{1}{3} + \frac{\cos k l}{(k l)^2} - \frac{\sin k l}{(k l)^3}\right] \left(\frac{F \epsilon^2 K_2(k \epsilon)}{2\mu}\right)^2.
\end{equation}
The energy content associated with the velocity field at a lengthscale $k^{-1}$ is given by
\begin{align}
E_k=4\pi k^2\langle \mathbf{\hat{U}}(\mathbf{k})\cdot\mathbf{\hat{U}}(\mathbf{-k})\rangle_{k=|\mathbf{k}|},
\label{eq:Ek}
\end{align}
yielding
\begin{align}\label{eq:Ek_dilute}
E_k = 4 \pi n \kappa^2 \left[ \frac{1}{3} + \frac{\cos(kl)}{(kl)^2} - \frac{\sin(kl)}{(kl)^3} \right] \frac{\epsilon^4 k^2}{l^2} K_2^2(k\epsilon).
\end{align}
In the double limit $\epsilon \rightarrow 0$ and $l \rightarrow 0$, this expression reduces to 
\begin{align}
E_k\approx \frac{8\pi}{15}n \kappa^2,
\end{align}
which is independent of the wavevector $k$.

\section{Results and discussion}

\begin{figure}[t]
\centering
  \includegraphics[height=7cm,angle=-90]{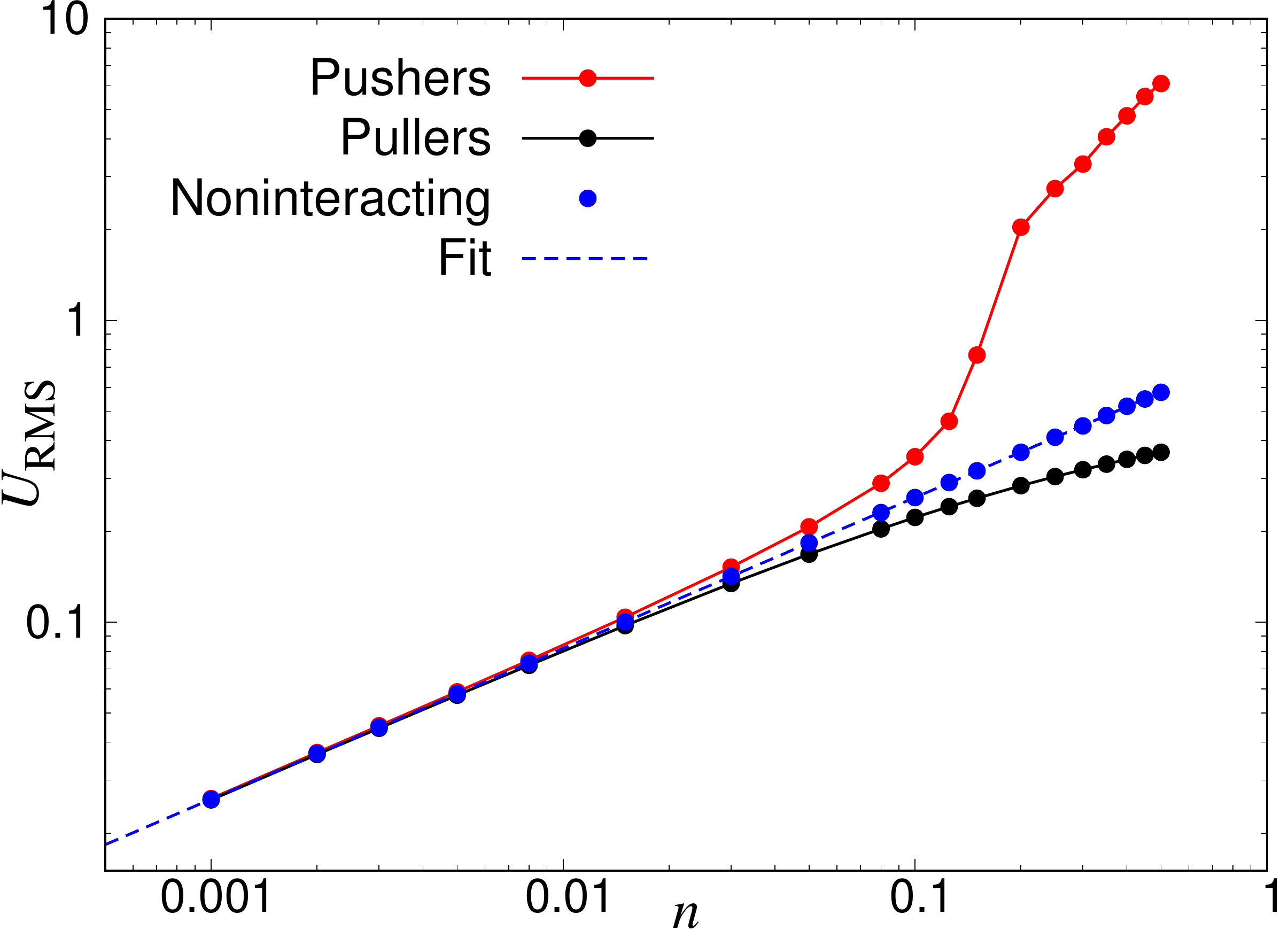}
  \caption{Root-mean-square fluid velocity $U_{\mathrm{RMS}}$, as a function of the swimmer number density $n$. The results were obtained using a linear box dimension $L=150$ for pushers and $L=100$ for noninteracting swimmers and pullers, due to the significant finite-size effects in the turbulent regime of the pusher suspensions. The dashed curve indicates a fit using Eq.~\eqref{eq:Usq}, yielding $\epsilon = 1.1$.}
  \label{fig:vrms}
\end{figure}

\subsection{Fluid statistics}

Figure \ref{fig:vrms} demonstrates how collective motion develops as a function of the concentration of microswimmers through the root-mean-square (RMS) fluid velocity $U_{\mathrm{RMS}} \equiv \langle \mathbf{U}^2\rangle ^{1/2}$ for suspensions of pushers, pullers and noninteracting swimmers; for the latter simulations the terms containing $\mathbf{U}$ in Eqs. \eqref{rdot} and \eqref{pdot} are set to zero, so that the swimmers do not interact with each other through the fluid. (Note again that, for the noninteracting case, pushers and pullers are equivalent.) For noninteracting swimmers, the RMS fluid velocity is accurately described by (the square root of) Eq.~\eqref{eq:Usq}, yielding an $n^{1/2}$ dependence over the whole concentration range. Using that expression, we fit the value $\epsilon \approx 1.1$ (while fixing all other parameters to the values from the LB simulations), in good accordance with the interpolation length of the Peskin $\delta$ function used in the LB simulations.~\cite{Peskin:2002} At intermediate concentrations for interacting swimmers, there is a deviation from the square root dependence, with the RMS velocity increasing faster than $n^{1/2}$ for pushers and slower for pullers: this is a signature of the build-up of long-ranged orientational correlations underlying the collective behaviour.\citep{Stenhammar1} At concentrations of $n > 0.2$ the turbulent state is fully developed for pushers, as is clearly visible in the snapshots in Fig. \ref{fig:snapshots} and videos available as ESI\dag. In the following sections, we will focus mainly on pusher suspensions and their transition from disordered swimming to active turbulence. 

\begin{figure}[t]
 \centering
 \includegraphics[height=11cm]{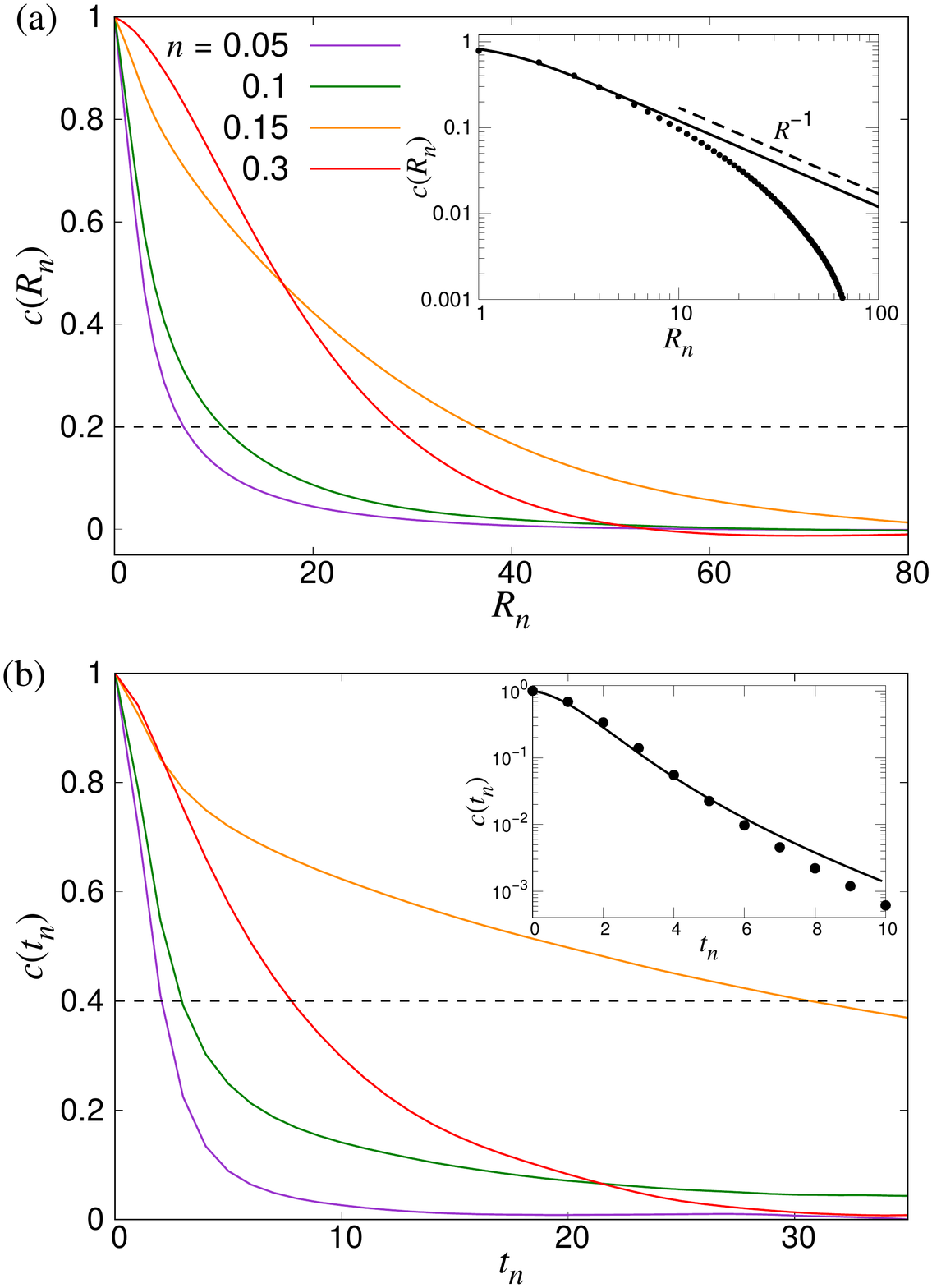}
 \caption{Build-up of long-ranged velocity correlations due to collective motion. (a) The spatial velocity correlation function $c(R_n)$, and (b) velocity autocorrelation functions $c(t_n)$ for four different densities as indicated. The dashed curve shows the value we used to determine the characteristic length- and timescales $\xi$ and $\tau$. The results were obtained using $L = 210$. Insets show comparisons between LB simulations of noninteracting swimmers (symbols) and the analytical predictions of Eqs.~\eqref{c_R_final} and \eqref{c_t_final}, using $\epsilon = 1.1$ (lines). }
 \label{fig:corr}
\end{figure}

\begin{figure}[t]
\centering
  \includegraphics[height=7cm,angle=-90]{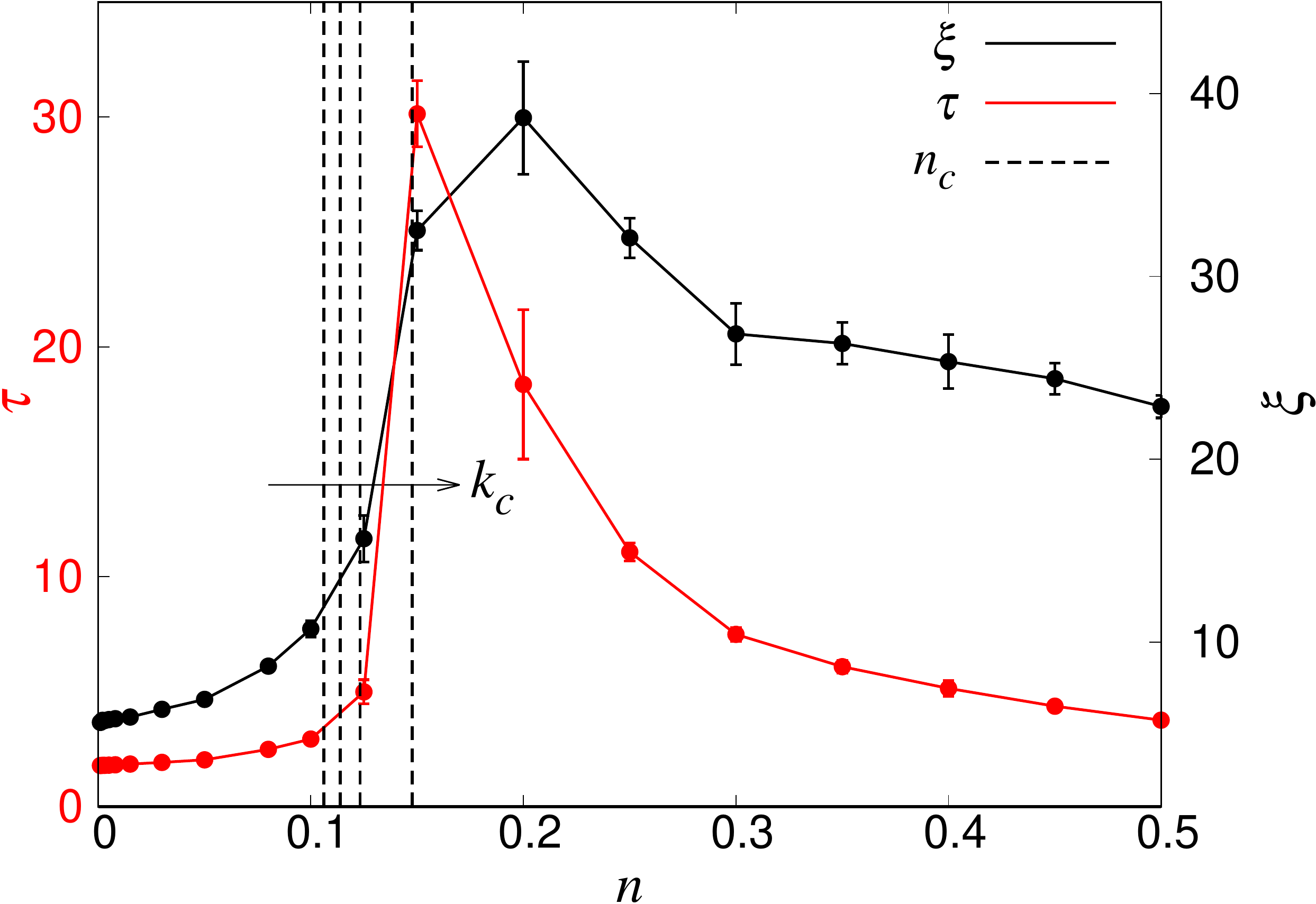}
  \caption{The characteristic length $\xi$ (black) and time $\tau$ (red) as a function of density $n$, obtained from systems with $L = 150$. Error bars indicate the estimated standard deviations obtained from dividing the simulation into four equally sized time intervals. The dashed curves indicate, from left to right, the predicted transition densities $n_c$ for an unbounded system ($k_c = 0$, Eq.~\eqref{rho_c}), and for the finite wavenumbers $k_c = 2\pi /L$, $4\pi/L$, and $8\pi / L$, as discussed in the text.}
  \label{fig:L150}
\end{figure}

To characterise the length- and timescales of the chaotic flows, we now turn to their spatial and temporal correlation functions. In Fig.~\ref{fig:corr}, we show both the equal-time velocity correlation functions and the velocity autocorrelation functions, computed for four different densities. While the temporal correlation function $c(t_n)$ in suspensions of noninteracting swimmers show reasonable agreement with the theoretical predictions (Eq.~\eqref{c_t_final}), the corresponding spatial curves start deviating from the predictions (Eq.~\eqref{c_R_final}) for $R_n \approx 10$, eventually falling below zero at $R_n \approx 80$. We attribute this poor agreement at intermediate and large $R_n$ to the significant effect of periodic boundary conditions on the long-ranged part of the flow fields as well as the presence of higher multipoles in the LB swimmer flow fields. All the correlation functions become increasingly long-ranged when going from $n = 0.05$ to $n = 0.10$. Their range, especially that of the autocorrelation function, then increases significantly near the transition around $n = 0.15$, followed by a slight decrease inside the turbulent regime ($n = 0.3$). To quantitatively characterise the length- and timescales of the chaotic flow, in accordance with Ref.~\cite{Thampi1}, we define the characteristic length $\xi$ as the distance where $c(R_n)$ has decreased to $0.2$ (Fig.~\ref{fig:corr}a), whereas the corresponding characteristic time $\tau$ is defined as the point when $c(t_n)$ has decayed to $0.4$ (see Figure \ref{fig:corr}b). The difference in the two threshold values is due to the slow decay of $c(t_n)$ around the transition density: it reaches $0.2$ only after prohibitively long times, resulting in poor statistics. The results, showing $\xi$ and $\tau$ as a function of $n$ for a wide range of densities are shown in Fig.~\ref{fig:L150}. In accordance with the results in Fig.~\ref{fig:corr}, there is a very sharp increase of both quantities at $n \approx 0.15$, followed by a gradual decrease towards a plateau value. The difference between the transition region and the turbulent region is most pronounced in the $\tau$ curve, while the maximum in $\xi$ is somewhat broader and has its peak at $n = 0.2$ rather than $n = 0.15$. The clearly non-monotonic curves reported here are different from the results reported previously by Saintillan and Shelley,\cite{Saintillan2} who did not observe any maximum at the transition density for neither the caracteristic length or time-scales, while the non-monotonic behaviour of $\tau(n)$ was previously observed by Krishnamurthy and Subramanian.~\cite{Krishnamurthy1} We attribute this difference to the significantly smaller system sizes used in earlier studies. 

We furthermore note that the predicted infinite-system critical density $n_c$ (Eq.~\eqref{rho_c}) falls somewhat below the observed increases in $\xi$ and $\tau$. In order to qualitatively explain this, we first consider the effect of a finite box size, which shifts the critical wavevector from $k_c = 0$ to $k_c = 2\pi / L$. In addition, the use of periodic boundary conditions will effectively screen the stresslet flow-fields, yielding them more short-ranged than the $r^{-2}$ spatial decay in an infinite fluid. While this screening is gradual, it becomes significant already at length scales between $L/4$ and $L/2$. Since the long-wavelength instability leading to active turbulence is an effect of the $r^{-2}$ decay of the flow field, this screening will shift the instability to even shorter lengthscales than the finite-box reasoning alone. Thus, in Fig.~\ref{fig:L150}, we also plot the values of $n_c$ corresponding to the critical wavenumbers $k_c = 0$, $2\pi/L$, $4\pi/L$ and $8\pi/L$, obtained through a linear stability analysis as detailed elsewhere.~\cite{Hohenegger1,Saintillan:PoF:2008} However, a more in-depth analysis of the effect of PBCs together with a more rigorous numerical treatment to numerically localise the transition would be necessary to confirm these qualitative arguments. 

\begin{figure}[t]
 \centering
 \includegraphics[height=9cm]{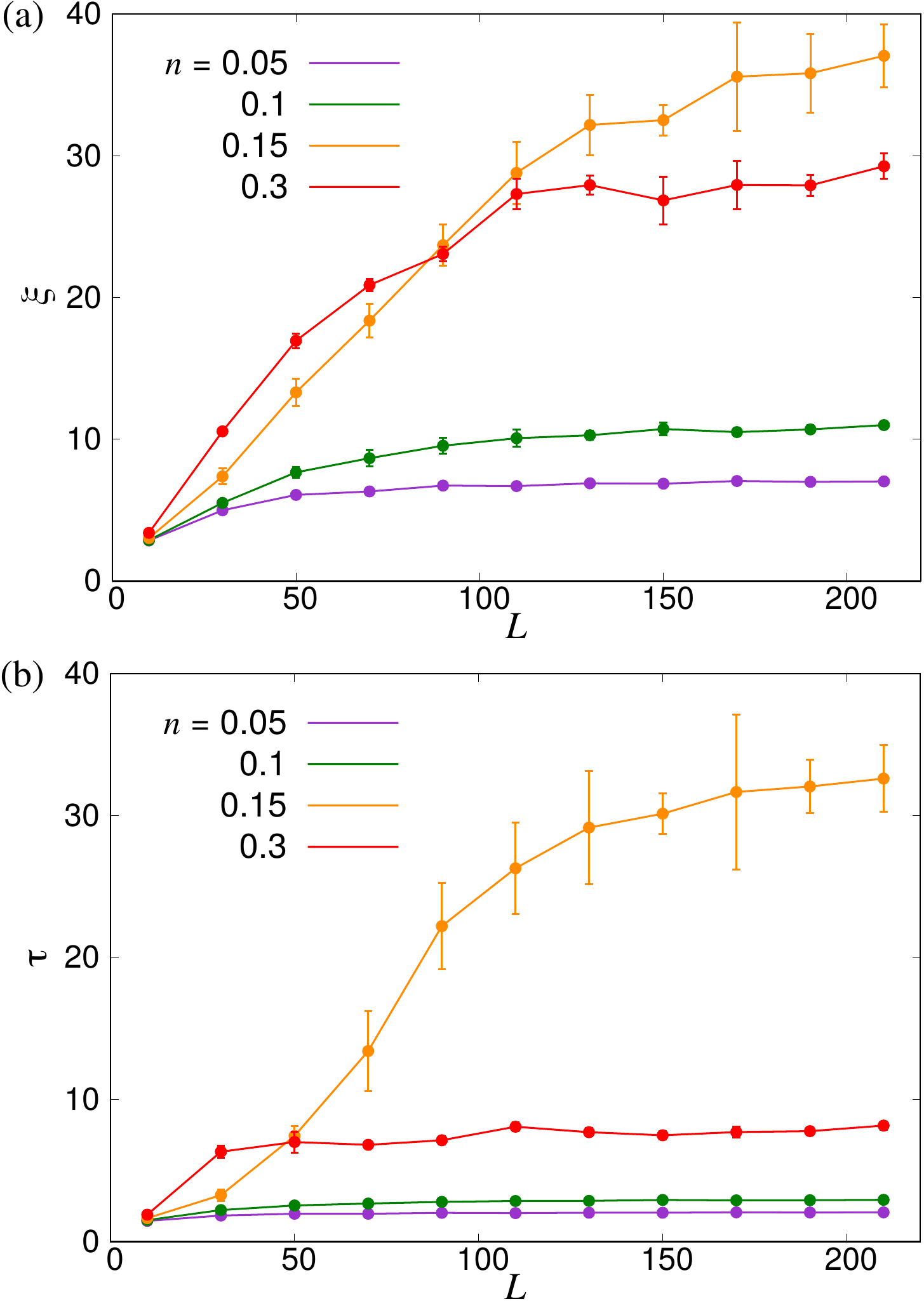}
 \caption{Finite-size effects of the characteristic length- and timescales. (a) $\xi$ and (b) $\tau$ plotted as functions of the linear system size $L$. Error bars indicate the estimated standard deviations obtained either from four independent runs with different initial conditions ($n = 0.15$) or from dividing the full simulation into four equally sized time intervals.}
 \label{fig:length_time}
\end{figure}

In order to further investigate the system size dependence discussed above, in Fig.~\ref{fig:length_time} we show $\xi$ and $\tau$ plotted as a function of the linear box size $L$ for the same four densities studied in Fig.~\ref{fig:corr}. In both panels, we observe significant finite-size effects until $L \approx 100$, corresponding to $N = 3 \times 10^5$ for $n = 0.3$, although these appear to persist to even larger systems in the transition region. These results again highlight the importance of using large-scale simulations when studying collective motion in microscopic models of microswimmers.

\begin{figure}[t]
\centering
  \includegraphics[height=8cm,angle=-90]{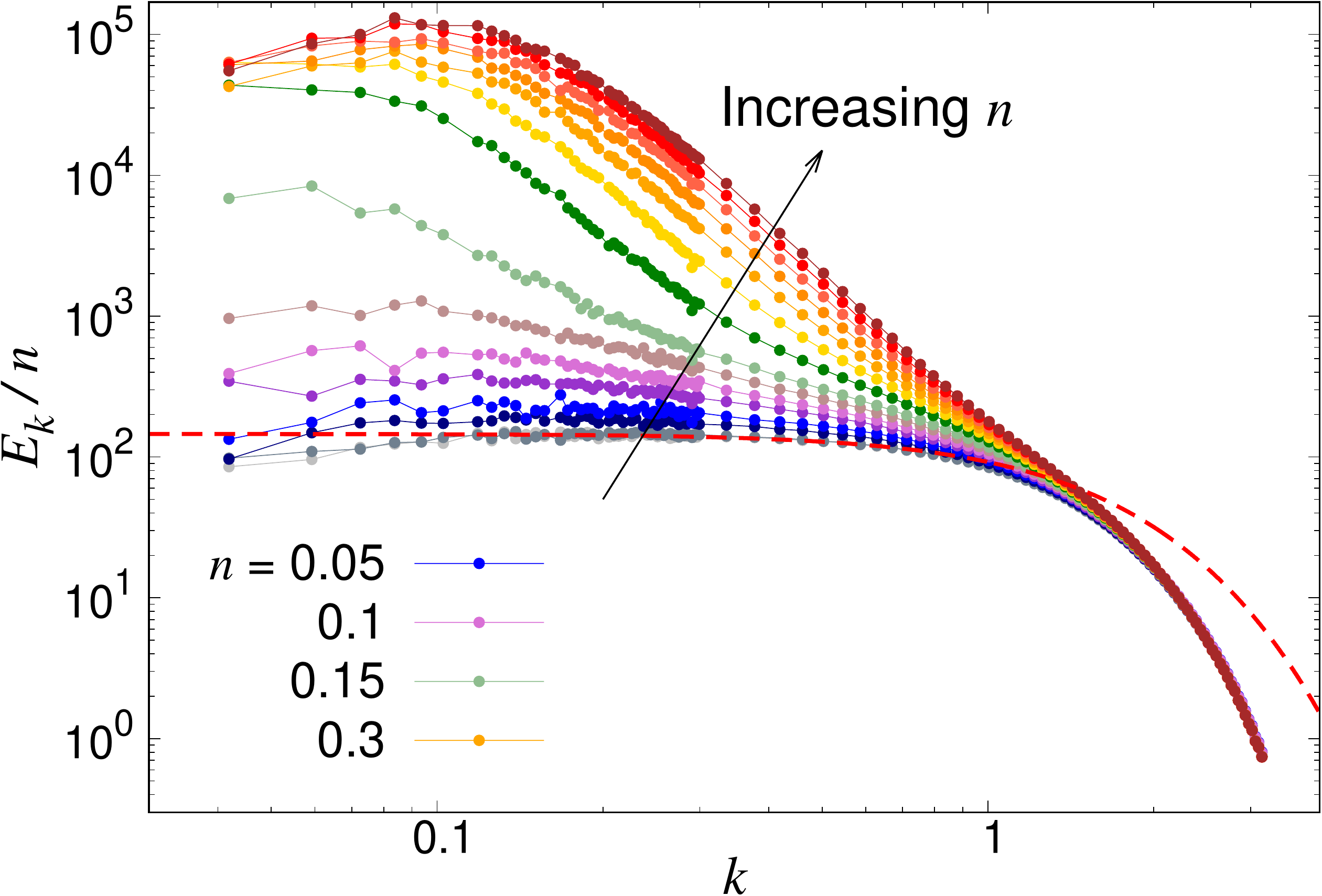}
  \caption{Energy spectra $E_k$, as defined in Eq.~\eqref{eq:Ek} for different pusher densities $n$: note the evolving peak for finite $k$ in the turbulent regime. Results were obtained using a system with $L=150$. The dashed line shows the low-density prediction of Eq.~\eqref{eq:Ek_dilute}, using the dialled value of $\kappa$ and fitted $\epsilon = 1.0$.}
  \label{fig:Ek}
\end{figure}

To further analyse the spatial structures of the flow, in Fig.~\ref{fig:Ek} we calculate the Fourier space energy spectrum $E_k$ as defined in Eq.~\eqref{eq:Ek}. For low densities, the spectrum (Fig.~\ref{fig:Ek}) is well described by the form predicted for uncorrelated swimmers in Eq.~\eqref{eq:Ek_dilute} with a flat shape at intermediate $k$, corresponding to a superposition of the $r^{-2}$ flow fields of uncorrelated swimmers. At the lowest accessible values of $k$, the spectra decrease slightly compared to the infinite-system prediction, again likely due to the effect of PBCs. At high $k$, corresponding to the length-scale of individual swimmers, $E_k$ decreases due to the short-range regularisation of the flow field; the slight discrepancy in between the data and the analytical prediction in this regime is due to the different forms of regularisation used in the two treatments. In the turbulent regime, most of the kinetic energy is localised at scales much bigger than $l$, in accordance with previous studies,\cite{Saintillan2,Krishnamurthy1}, even though the collective motion is driven by energy injected at small lengthscales. Another feature of the energy spectrum, which was absent in previous studies due to finite-size effects, is the peak in the spectrum which evolves for high swimmer densities, again indicating a characteristic, finite length-scale of the flow field. This should be contrasted with the curve corresponding to the transition density $n = 0.15$ (light green curve in Fig.~\ref{fig:Ek}), which monotonically increases as $k \rightarrow 0$.

\begin{figure}[t]
\centering
  \includegraphics[height=7cm,angle=-90]{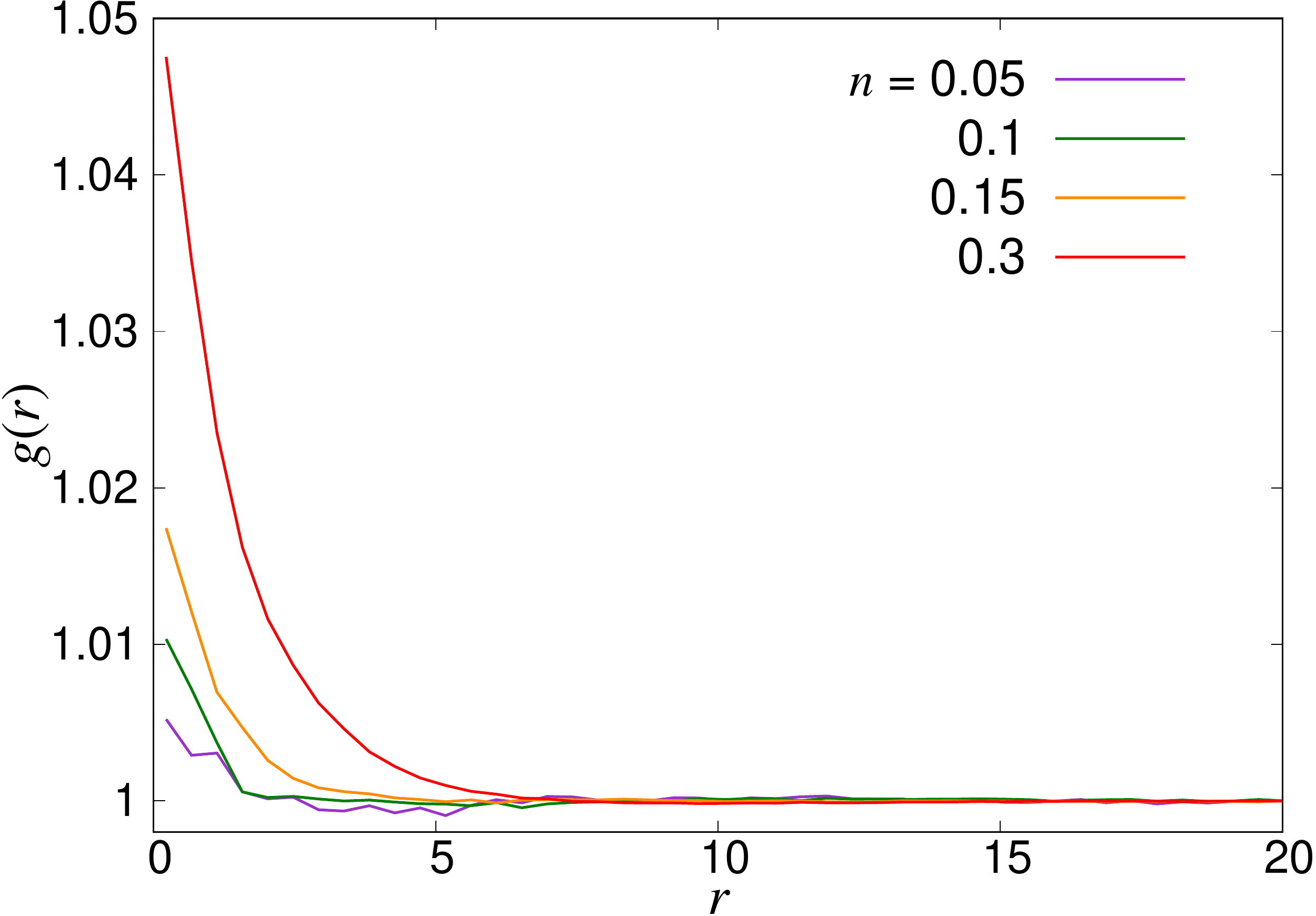}
  \caption{Radial distribution function $g(r)$ for pusher suspensions of four different densities $n$; for pullers, the corresponding curves (not shown) are completely flat at all densities within the statistical uncertainty.}
  \label{fig:rdf}
\end{figure}

\subsection{Swimmer statistics}

\begin{figure}[t]
\centering
\includegraphics[height=5.5cm]{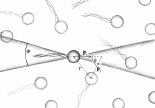}
\caption{Schematic image of the orientational correlation functions, where $\mathbf{p}$ is the orientation of a swimmer, showing the angles $\theta$ and $\varphi$ used to sample the order parameters $P(r)$ and $S(r)$. }
\label{fig:orientation}
\end{figure}

\begin{figure}[p]
 \centering
 \includegraphics[height=10cm]{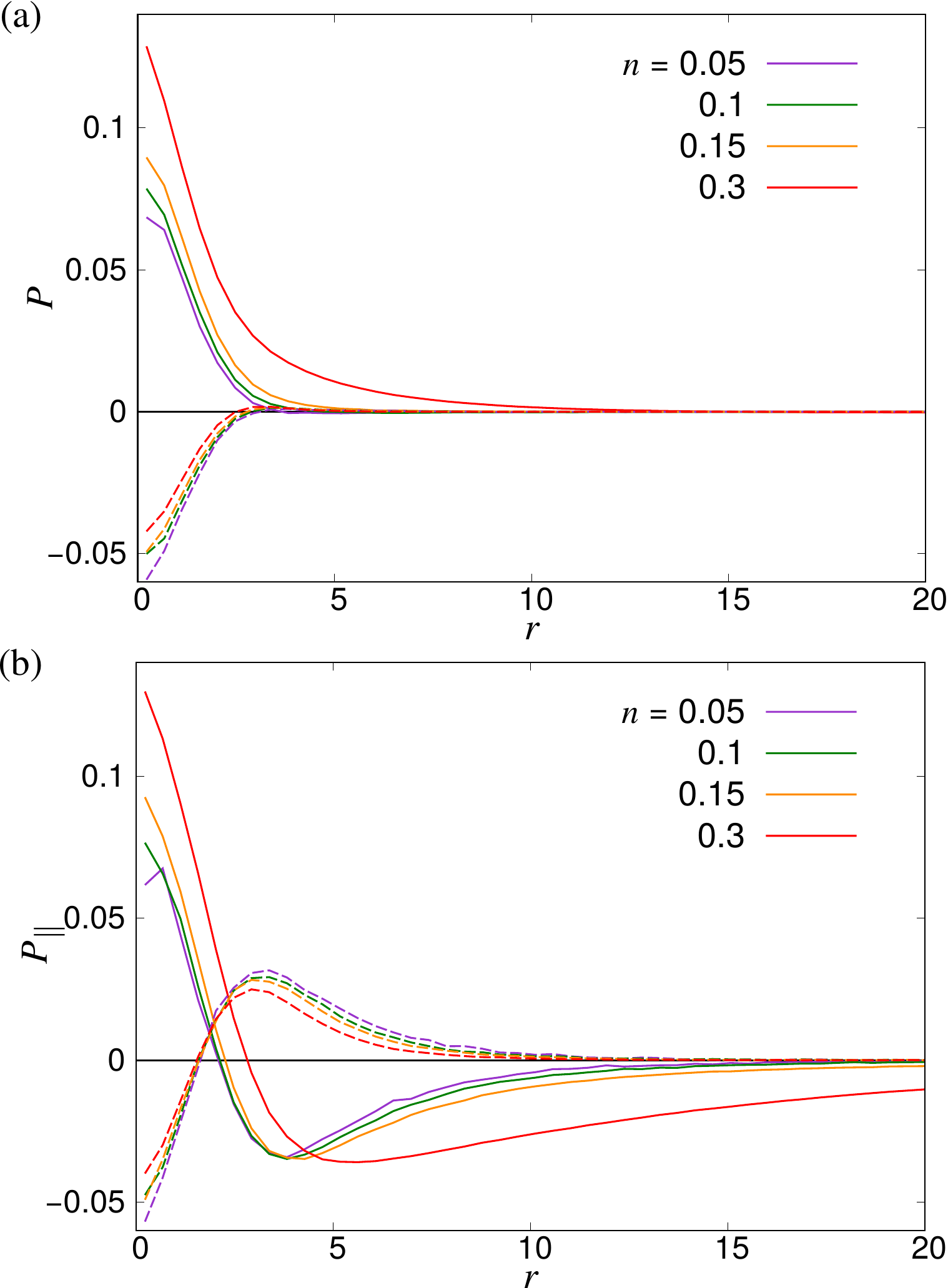}
 \caption{Distance-dependent polar order parameter $P(r)$ as defined in Eq.~\eqref{P_r} (a) sampled isotropically, and (b) sampled in a cone of angle $\varphi = 20$ degrees in front of and behind the swimmer as shown in Fig.~\ref{fig:orientation}. Solid curves denote pushers and dashed curves pullers. The apparent kinks at small $r$ are due to poor sampling at small separations. }
 \label{fig:p}
\end{figure}

\begin{figure}[p]
 \centering
 \includegraphics[height=7cm,angle=-90]{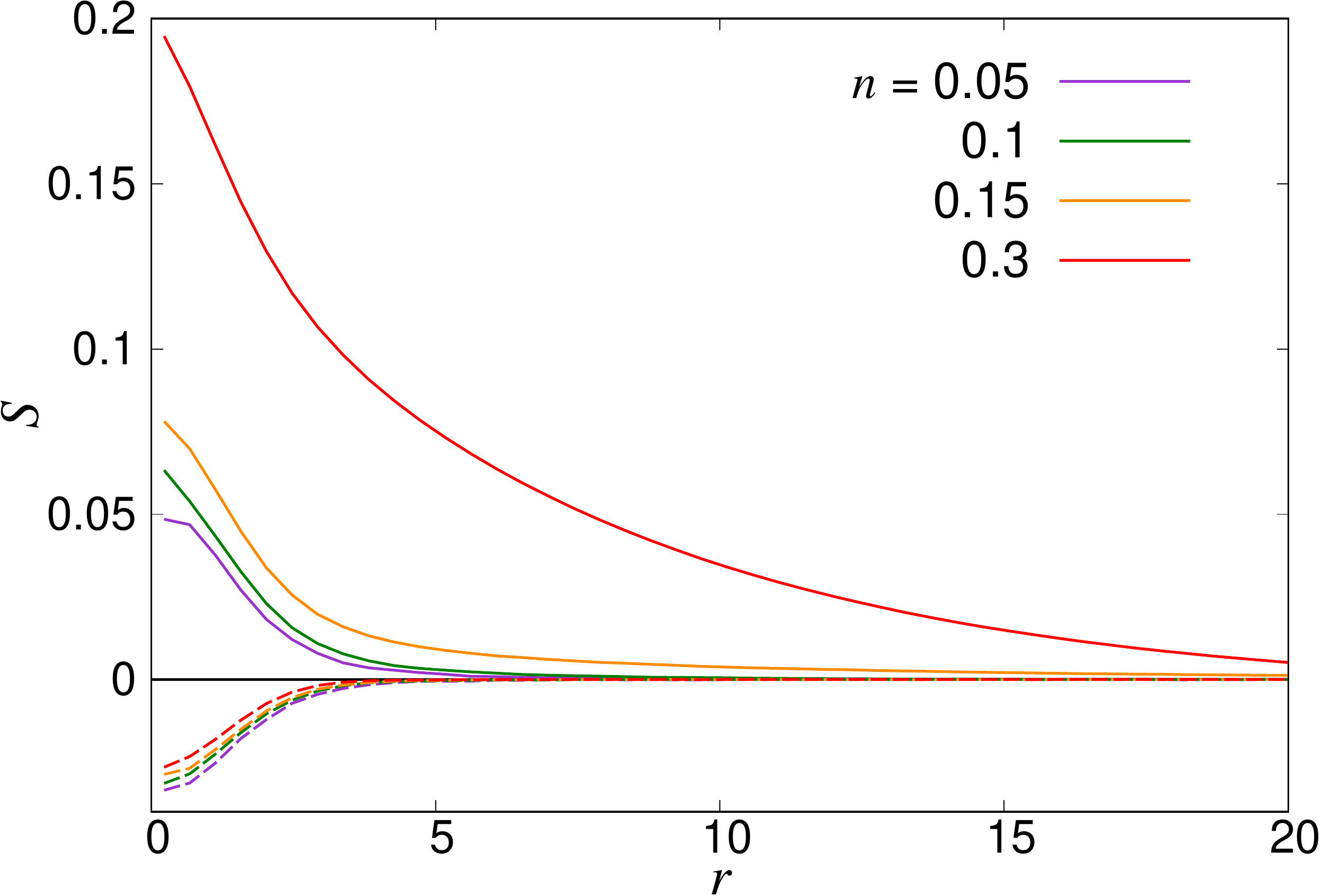}
 \caption{Distance-dependent nematic order parameter $S(r)$ as defined in Eq.~\eqref{S_r} for pushers (solid curves) and pullers (dashed curves).}
 \label{fig:s}
\end{figure}

In order to further characterise the properties of the suspension we now turn to the local ordering of the \emph{swimmers}. First of all, there are no significant density inhomogeneities in any of the systems: the pair correlation function $g(r)$ (Fig.~\ref{fig:rdf}) has a maximum peak height of $\sim 1.05$ inside the turbulent regime. This is in accordance with previous results~\cite{Saintillan1,Subramanian1,Stenhammar1,Saintillan4,Hohenegger1} showing that the transition to collective motion is orientational in nature. We thus turn to analyse the orientational order between swimmers, as quantified by the angle $\theta$ between the orientations of two swimmers separated by a distance $r$ (Fig.~\ref{fig:orientation}). The polar and nematic order parameters were calculated as a function of swimmer-swimmer separation in two different ways: first, for all swimmers a distance $r$ from the central swimmer (solid curves in Fig. \ref{fig:p}a and \ref{fig:s}), and secondly, for only those swimmers that lie along the axis of the swimming direction $\mathbf{p}$ of the central swimmer, \emph{i.e.} within two cones centered around $\theta=0$ and $\theta=\pi$. The polar and nematic order parameters $P(r)$ and $S(r)$ are furthermore defined as
\begin{equation}\label{P_r}
P(r) = \langle P_1(\cos \theta)\rangle_{|\mathbf{r}_i - \mathbf{r}_j| = r} = \langle \cos \theta \rangle_{r}
\end{equation}
and
\begin{equation}\label{S_r}
S(r) = \langle P_2 (\cos \theta)\rangle_{|\mathbf{r}_i - \mathbf{r}_j| = r} = \Bigg\langle \frac{3\cos^2 \theta -1}{2}\Bigg\rangle_{r}
\end{equation}
where $P_1$ and $P_2$ are the first and second Legendre polynomials. All the order parameters were calculated both for pushers (solid curves) and pullers (dashed curves). Looking at the full polar order parameter $P(r)$ (Fig.~\ref{fig:p}a), we observe a weak local polar alignment for pushers and antialignment for pullers, which however converge rather quickly to zero around $r \approx 5$. In the case where the polar order parameter was calculated along $\mathbf{p}$ (Fig.~\ref{fig:p}b), the curves for pushers instead fall below zero, showing that the swimmers are weakly aligned in \emph{opposite} directions along their swimming direction for $r \geq 3$. Looking at the corresponding $S(r)$ curves in Fig.~\ref{fig:s}, for pushers (solid curves) we observe a significantly more pronounced increase in both the magnitude and range of the order parameter when going into the turbulent state. Furthermore, unlike the case of $P(r)$, no significant anisotropy in the nematic order parameter is observed (data for $S_{||}$ not shown). These observations indicate that far-field hydrodynamics alone is sufficient to induce significant nematic ordering between swimmers, while the effect of the polar symmetry breaking due to self-propulsion is subdominant. 

\begin{figure}[h]
 \centering
 \includegraphics[height=10cm]{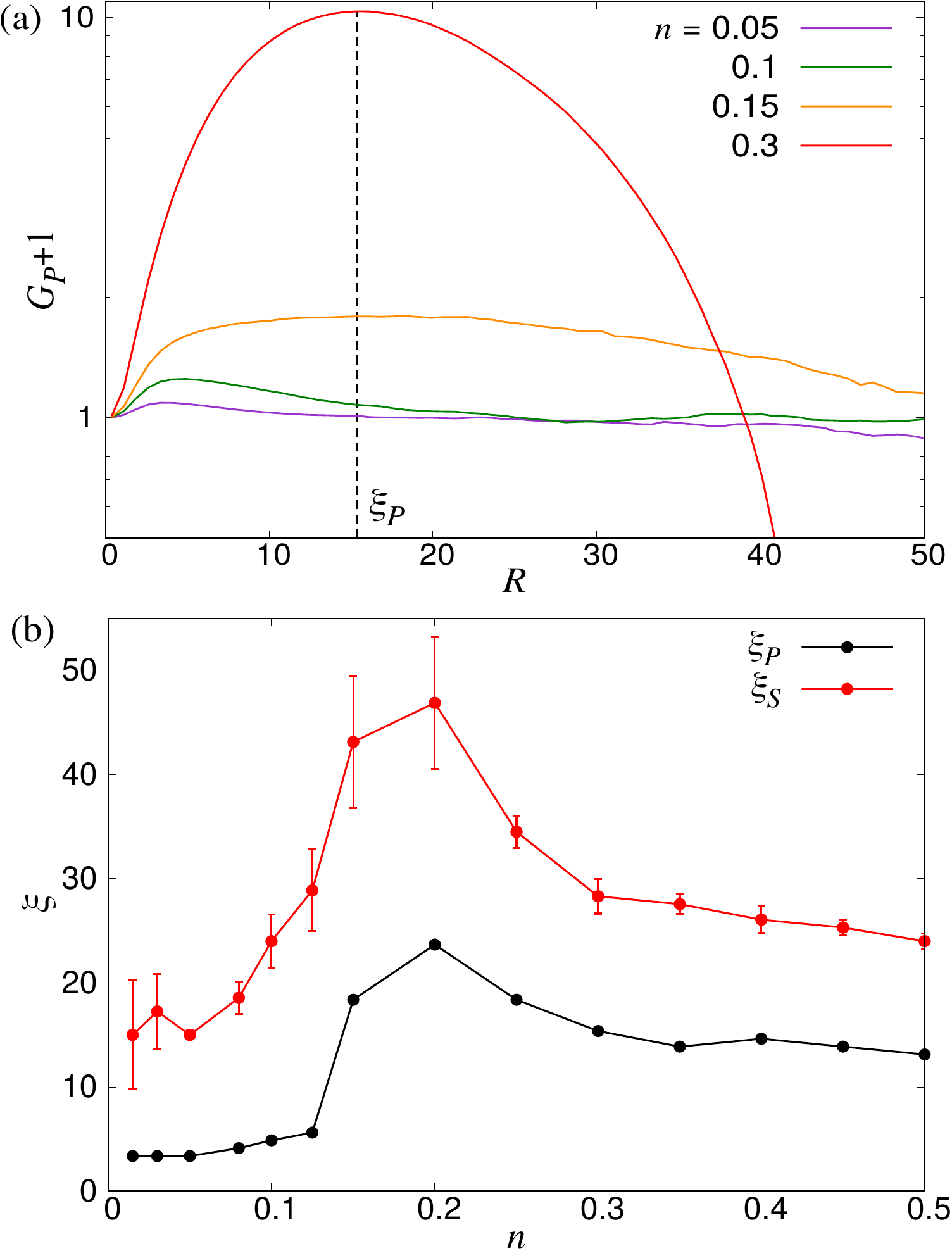}
 \caption{Characteristic lengthscales as measured from the orientational order between swimmers. (a) Cumulative polar order parameter $G_P$, as defined in Eq.~\eqref{xi_P}, together with the definition of the corresponding lengthscale $\xi_P$; $\xi_S$ is defined analogously. The $y$-axis has been shifted by unity for visualisation purposes. (b) $\xi_P$ and $\xi_S$ as a function of the microswimmer density $n$: note the similarity between $\xi_S$ and $\xi$ as measured from the fluid velocity correlations (Fig.~\ref{fig:L150}). Error bars indicate the estimated standard deviations obtained dividing the simulation into four equally sized time intervals.}
 \label{fig:kirkwood}
\end{figure}

Another approach to calculate the characteristic lengthscales based on the polar and nematic order parameters of the swimmers is to consider the integrated form of the order parameters $P(r)$ and $S(r)$ in Eqs. \eqref{P_r}--\eqref{S_r}, \emph{i.e.}
\begin{eqnarray}
& G_P (R) \equiv \displaystyle \int_0^R P(r) 4\pi n r^2 dr,\label{xi_P} \\
& G_S (R) \equiv \displaystyle \int_0^R S(r) 4\pi n r^2 dr.\label{xi_S}
\end{eqnarray}
These forms of the order parameter measure the range of polar and nematic order around a single swimmer, in a manner equivalent (modulo a constant shift of unity) to the distance-dependent Kirkwood $G$-factor employed to measure local order in polar fluids.~\cite{Bottcher1} Based on these order parameters, we define the characteristic lengthscales $\xi_P$ and $\xi_S$ as the values of $R$ where $G_P(R)$ and $G_S(R)$ take on their maximum values (see Fig.~\ref{fig:kirkwood}a). The resulting lengthscale curves are shown in Fig.~\ref{fig:kirkwood}b as a function of density. A direct comparison between $\xi_P$ and the characteristic lengthscale calculated from the fluid (Fig.~\ref{fig:L150}) shows a striking similarity between the curves, modulo a shift in the $y$ direction attributable to the somewhat arbitrary cutoff value used to calculate $\xi$ from the fluid flows. The curve for $\xi_S$ shows a similar non-monotonic shape as the two other curves. It is, however, shifted towards slightly higher values of $\xi$ compared to the $\xi_P$ curve, in accordance with the observations made in Figs.~\ref{fig:p} and~\ref{fig:s}, and exhibits somewhat larger statistical fluctuations. Taken together, however, our three separate analyses (Figs.~\ref{fig:length_time} and~\ref{fig:kirkwood}b) of the emerging lengthscales based on either fluid flows or the swimmer orientation provides a consistent picture showing a sharply increasing $\xi$ near the transition, which then plateaus to a finite value in the turbulent regime. 

\section{Conclusions}

In this study, we have provided an analysis of the structure and dynamics in suspensions of swimming microorganisms using lattice Boltzmann simulations of model microswimmers. The model, which accurately includes the effect of far-field stresslet flow fields, enables us to simulate large enough systems ($N \approx 3 \times 10^6$) to quantitatively study the spatio-temporal properties of the ensuing long-range fluid flows in the turbulent regime, something which has not been possible using more complex models due to the high computational costs. For non-interacting swimmers, we tested the method against a number of analytical predictions for the structural and dynamical observables. In the semidilute regime, where swimmer-swimmer interactions become significant, we showed that the correlation length $\xi$ and correlation time $\tau$ of the flows diverge steeply near a swimmer density close to the one predicted from kinetic theory, after a qualitative inclusion of the effect of periodic boundary conditions. Beyond the transition, both these quantities relax to significantly smaller values, indicating the emergence of flows with a finite, well-defined length- and timescale. This was further confirmed in the Fourier-space energy spectra of the flow field, which shows the development of a peak at low $k$ in the turbulent regime. The statistics of the swimmer-swimmer correlations are consistent with the above analysis: the density-dependent lengthscale $\xi_P$ derived from the local polar ordering between swimmers matches qualitatively that calculated from the flow field, while the corresponding lengthscale $\xi_S$ calculated from the nematic order parameter is significantly larger than $\xi_P$, albeit with a similar shape, indicating a longer range of the local nematic order due to hydrodynamic interactions. 

Apart from the results described above, our study highlights the need for employing very large systems when studying collective behaviours in microswimmer suspensions: our finite-size studies indicate that linear system sizes of at least 100 times the swimmer length is necessary to capture the properties of the chaotic flows. This computational efficiency comes at the cost of ignoring effects of near-field hydrodynamic and short-ranged steric interactions between microswimmers, which will become significant at high microswimmer densities. On the other hand, the simplicity of the model provides us with full control of how the different system parameters affect the collective behaviour, and enables a direct comparison with kinetic theories that employ microswimmer models with long-range stresslet flow fields.~\cite{Stenhammar1} The model can also be extended to include the effect of external gradients, system boundaries, and short-ranged interactions, thus providing further insight into the interplay between microscopic interactions and dynamics, external perturbations, and collective behaviour. 

\section*{Conflicts of interest}
There are no conflicts to declare.

\section*{Acknowledgements}
Joost de Graaf, Davide Marenduzzo and Rupert Nash are kindly acknowledged for helpful discussions during the early phases of this project. The work was funded (DB, JS) by the Swedish Research Council (grant ID 2015-05449) and the Crafoord Foundation (grant ID 20170678). CN acknowledges the support of an Aide Investissements d'Avenir du LabEx PALM
(ANR-10-LABX-0039-PALM). All simulations were performed on resources provided by the Swedish National Infrastructure for Computing (SNIC) at LUNARC. 


\balance


\bibliography{Ref1} 
\bibliographystyle{rsc} 

\end{document}